\documentclass[aps,twocolumn,pre,superscriptaddress,normalem]{revtex4-2}
\usepackage{xcolor,hyperref}
\hypersetup{
   colorlinks,
   linkcolor={blue!50!black},
   citecolor={blue!50!black},
   urlcolor={blue!80!black}
}

\usepackage{amssymb}

\usepackage{epsfig, amsmath}
\usepackage{bm}
\usepackage{soul,ulem}
\usepackage{hyperref}
\usepackage{footmisc}
\usepackage{wrapfig}

\DeclareMathAlphabet{\mathitbf}{OML}{cmm}{b}{it}

\newcommand{\xv}{\mathitbf x}
\newcommand{\piv}{\bm{\pi}}
\newcommand{\psiv}{\bm{\psi}}
\newcommand{\calBold}[1]{\mbox{\boldmath${\cal #1}$}}

\newcommand{\dbar}{{\,\mathchar'26\mkern-12mu d}}

\newcommand{\sFrac}[2]{{\textstyle\frac{#1}{#2}}}
\newcommand{\tripleCdot}{:\!\cdot\,}

\usepackage{graphicx}
\begin{document}

\title{Disordered crystals reveal soft quasilocalized glassy excitations}
\author{E.~Lerner}
\email{e.lerner@uva.nl}
\affiliation{Institute for Theoretical Physics, University of Amsterdam, Science Park 904, Amsterdam, Netherlands}
\author{E.~Bouchbinder}
\email{eran.bouchbinder@weizmann.ac.il}
\affiliation{Chemical and Biological Physics Department, Weizmann Institute of Science, Rehovot 76001, Israel}

\begin{abstract}
Structural glasses formed by quenching a melt are known to host a population of low-energy quasilocalized (nonphononic) excitations whose frequencies $\omega$ follow a universal $\sim\!\omega^4$ distribution as $\omega\!\to\!0$, independently of the glass formation history, the interparticle interaction potential or spatial dimension. Here, we show that the universal quartic law of nonphononic excitations also holds in disordered crystals featuring finite long-range order, which is absent in their glassy counterparts. We thus establish that the degree of universality of the quartic law extends beyond structural glasses quenched from a melt. We further find that disordered crystals, whose level of disorder can be continuously controlled, host many more quasilocalized excitations than expected based on their degree of mechanical disorder --- quantified by the relative fluctuations of the shear modulus --- as compared to structural glasses featuring a similar degree of mechanical disorder. Finally, we show that the stability bound on nonlinear quasilocalized excitations --- previously established in structural glasses --- also holds in disordered crystals. Our results are related to glass-like anomalies experimentally observed in disordered crystals. More broadly, they constitute an important step towards tracing the essential ingredients necessary for the emergence of universal nonphononic excitations in disordered solids.
\end{abstract}

\maketitle

\section{Introduction}
It is now well established that structural glasses formed by quenching a melt through the glass transition temperature generically feature a population of soft, quasilocalized (nonphononic) excitations (QLEs)~\cite{JCP_Perspective}. These excitations are expected to play key roles in determining various static, mechanical, transport and dynamic properties of glasses, and possibly of some supercooled liquids as well~\cite{harrowell_soft_modes_supercooled_liquids_nature_2008,harrowell_2009,jeppe_project_jcp,Sylvain_stzs_in_supercooled_liquids2021}. QLEs have been envisioned, and their importance has been underlined, since the late 1980s by several workers, as described in detail in a recent review~\cite{JCP_Perspective}. A first-principle understanding of the emergence of low-energy quasilocalized excitations in structural glasses is still incomplete, despite some recent progress in identifying and solving potentially relevant mean-field models~\cite{scipost_mean_field_qles_2021,meanfield_qle_pierfrancesco_prb_2021,pierfrancesco_arXiv_2021,vector_spin_glasses_scipost_2022}.

Intriguingly, the vibrational density of states (VDoS) ${\cal D}(\omega)$ of low-energy QLEs --- of vibrational frequency $\omega$ --- has been shown to follow a universal law in structural glasses, of the form ${\cal D}(\omega)\!\sim\!\omega^4$, independently of the interparticle interaction potential~\cite{modes_prl_2020,sticky_spheres1_karina_pre2021}, spatial dimension~\cite{modes_prl_2018} or the glass formation protocol~\cite{modes_prl_2016,pinching_pnas,LB_modes_2019}. Does this nonphononic quartic law remain valid in a broader class of disordered solids, beyond structural glasses formed by quenching a melt? Can it be observed in disordered solids featuring some long-range order? In this work, we address these questions by studying the statistical, structural and mechanical properties of soft excitations that emerge in a class of \emph{disordered crystals}. To this aim, we study a variant of a model system put forward by Barrat and coworkers~\cite{barrat_2013_amorphization,barrat_2014_pnas_amorphization,barrat_2016_prb_amorphization} in which microscopic disorder can be gradually introduced into an initially perfectly-crystalline solid, until the occurrence of a global amorphization transition of the disordered crystal into a glassy state. Similar procedures were carried out in two-dimensional systems in~\cite{Tong2015,D0SM01989E,ning_xu_2022}.

We employ a variant of these aforementioned three-dimensional disordered crystal models in which the external pressure is fixed under variations of the microscopic disorder, controlled by a dimensionless parameter $\delta\!\in\![0,1]$ ($\delta\!=\!0$ corresponds to a perfect crystal). Additional details about the model can be found below and in Appendix~\ref{app:model}. After identifying the amorphization transition point, we study the vibrational spectra of disordered crystals \emph{prior} to amorphization, i.e.~in states that feature finite long-range crystalline order. We find that these disordered crystals host a population of QLEs whose frequencies obey the universal $\sim\!\omega^4$ nonphononic law. We thus establish that the degree of universality of the nonphononic quartic law of QLEs in disordered media extends beyond structural glasses quenched from a melt. In addition, our results shed new light on low-temperature, glass-like thermodynamic, transport and vibrational anomalies observed in laboratory disordered crystals~\cite{ackerman1981glassy,Anderson_1985, Pohl_disordered_crystals_1992,ramos1997quantitative,pohl_review,ramos2003low,vdovichenko2015thermal,gebbia2017glassy,moratalla2019emergence,Ramos_2020}.

\begin{figure*}[ht]
\centering
\begin{tabular}{ccc}
\includegraphics[width=1.0\linewidth]{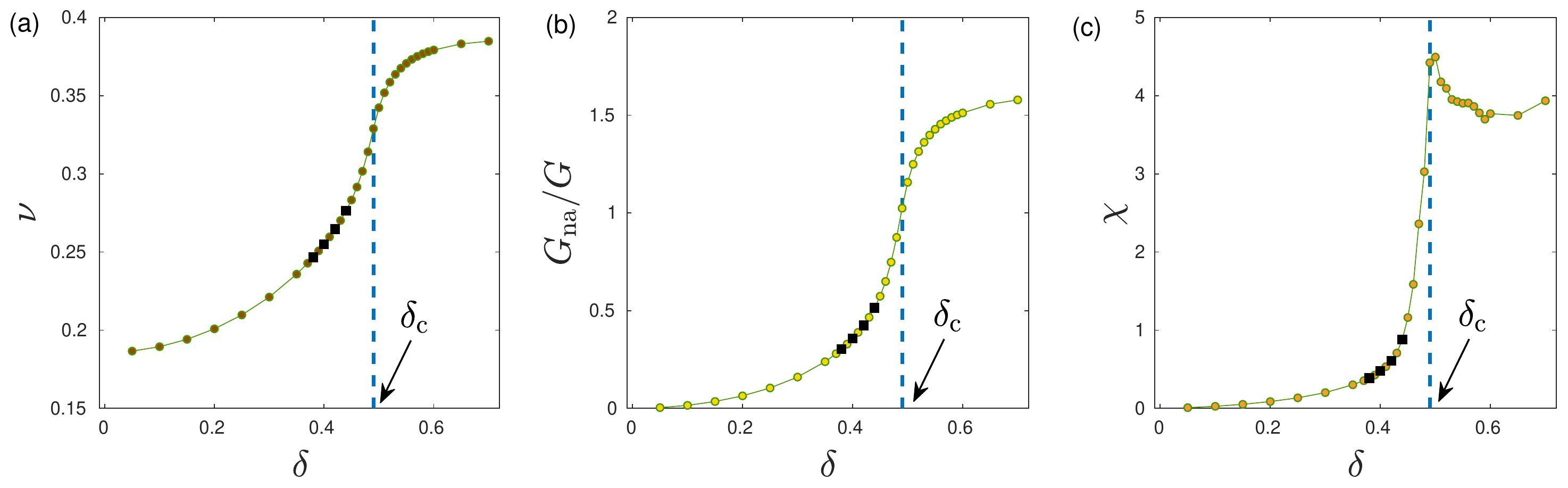}
\end{tabular}
\caption{Dimensionless elastic properties across the amorphization transition: (a) Poisson's ratio $\nu$, (b) nonaffine-to-affine shear moduli ratio $G_{\rm na}/G$, (c) relative sample-to-sample fluctuations of the shear modulus $\chi$, see precise definitions in Appendix~\ref{app:observables}. These data were measured over about 700 independent realizations of systems of $N\!=\!108000$ particles at fixed vanishing pressure. The vertical lines mark an estimation (based on $\chi$) of the transition point at $\delta_{\rm c}\!\approx\!0.49$. The black squares mark the disordered-crystal ensembles studied below, namely $\delta\!=\!0.38,0.40,0.42,0.44$, all of which are located below the amorphization transition.}
\label{fig:fig1}
\end{figure*}

We then go on to assess the degree of mechanical disorder of these disordered crystals, and compare their relative abundance of QLEs to that found in structural glasses featuring similar degrees of mechanical disorder. This analysis reveals --- surprisingly --- that these disordered crystals host many more QLEs as compared to structural glasses. Finally, we study the micromechanical properties of QLEs in disordered crystals, e.g.~the degree of their spatial localization and stability bounds, and find striking similarities to the corresponding QLEs' properties in structural glasses. The micromechanical properties of QLEs in disordered crystals also reveal signatures of the proximity to the amorphization transition.

\section{Disordered crystals model and the amorphization transition}

We employ a binary system of $N$ Lennard-Jones particles amongst which a fraction $x$ is of one species (`small' particles) and the remaining particles are of another species (`large' particles). One species may be regarded as `impurity atoms' relative to the other species (`host atoms'), as is extensively used in laboratory disordered crystals. The parameter $\delta$ determines the effective interaction of the two species, such that $\delta\!=\!0$ corresponds to a mono-disperse system, while $\delta\!=\!1$ corresponds to the largest-considered interaction contrast between the two species. Details about the pairwise interaction potential and system sizes employed are provided in Appendix~\ref{app:model}. Throughout this study we fix $x\!=\!1/2$, such that $\delta\!=\!1$ corresponds to a 50:50 binary mixture extensively employed in computer glass-forming models. We have verified that the key results reported below do not qualitatively depend on the precise choice of $x$.

\begin{figure}[ht]
\centering
\begin{tabular}{ccc}
\includegraphics[width=1.0\linewidth]{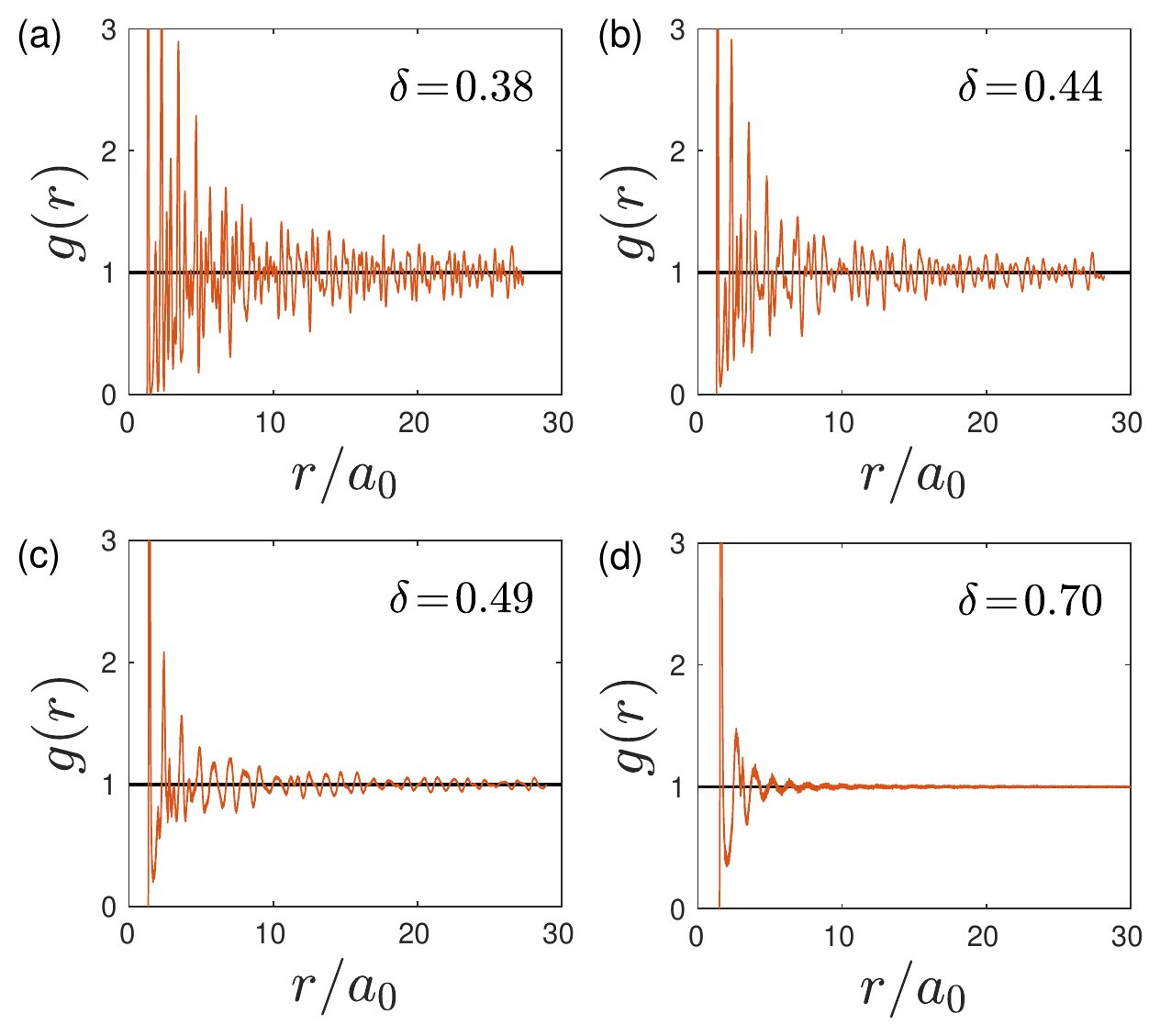}
\end{tabular}
\caption{Pair correlation functions $g(r)$ between the `large' particle-species (see Appendix~\ref{app:model} for details), measured for ensembles of disordered solids of $N\!=\!108000$ particles created with different $\delta$ values, as indicated in the legends. We indeed find that all the way up to the amorphization transition at $\delta_{\rm c}\!\approx\!0.49$, long-range order persists.}
\label{fig:fig2}
\end{figure}

Disordered crystals are generated by first placing the two particle species randomly on a perfect fcc lattice, and setting $\delta\!=\!0$. Then, for each value of the parameter $\delta\!>\!0$, the potential energy is relaxed under fixed zero confining pressure to obtain a disordered crystal. This energy relaxation step spontaneously generates positional disorder and internal stresses, absent in the reference fcc lattice. Continuously varying $\delta$ gives rise to a class of disordered crystals with a varying degree of disorder. Once $\delta$ exceeds a critical value --- to be determined next ---, the disordered crystal collapses into a fully disordered, glassy state via a so-called `amorphization transition'~\cite{barrat_2013_amorphization,barrat_2014_pnas_amorphization,barrat_2016_prb_amorphization}.

\begin{figure*}[ht]
\centering
\begin{tabular}{ccc}
\includegraphics[width=1.00\linewidth]{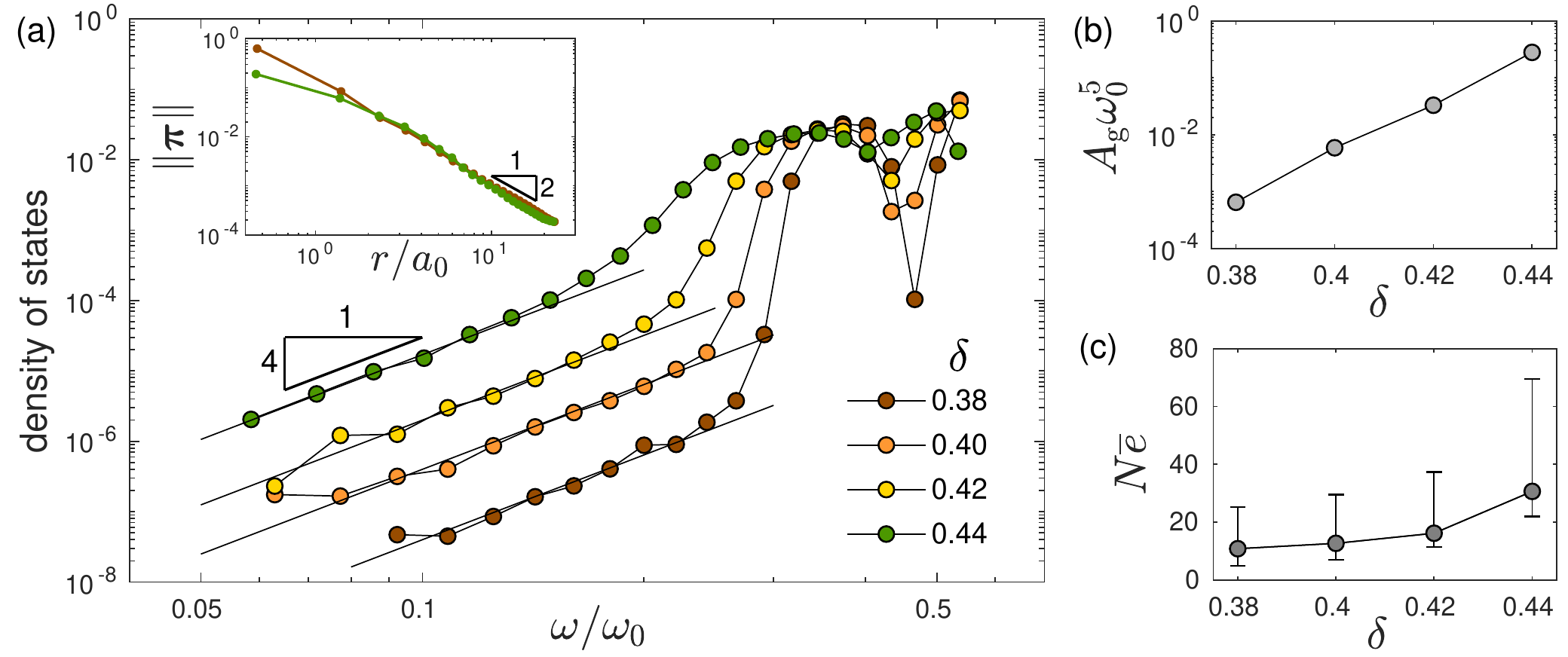}
\end{tabular}
\caption{(a) The vibrational density of states of disordered crystals, for different values of the parameter $\delta$ (as indicated in the legend), plotted against rescaled frequency $\omega/\omega_0$, where $\omega_0\!\equiv\!c_{\rm s}/a_0$ with $c_{\rm s}$ being the shear wave-speed and $a_0\!\equiv\!(V/N)^{1/3}$ is an interparticle distance. The continuous lines correspond to fits to the $\sim\!\omega^4$ law, from which the prefactors $A_{\rm g}$ are extracted. Inset: the spatial decay of (two randomly selected) soft quasilocalized excitations $\piv$ (obtained by the nonlinear framework presented in Appendix~\ref{app:observables}) follows the expected continuum-like $\sim\!r^{-2}$ scaling in three dimensions; the colors match the $\delta$ values in the legend. (b) Dimensionless prefactors $A_{\rm g}\omega_0^5$ vs.~$\delta$ on a log-lin scale. (c) The products $N\bar{e}$ --- representing the mean size of the disordered cores of QLEs ($e$ is the participation ratio, defined in the text) --- plotted vs.~$\delta$. The vertical bars cover the 2nd and 3rd quartiles, see text for additional discussion.}
\label{fig:fig3}
\end{figure*}

To quantitatively identify this transition, we study the behavior of three dimensionless observables while varying the control parameter $\delta$: Poisson's ratio $\nu$, the ratio of the nonaffine-to-affine shear modulus $G_{\rm na}/G$, and the relative fluctuations of the shear modulus $\chi$. Figure~\ref{fig:fig1} displays our results, see figure caption for details. It is observed that while $\nu$ and $G_{\rm na}/G$ vary smoothly and monotonically with $\delta$, $\chi(\delta)$ exhibits a rather sharp peak, which enables the identification of the location of the amorphization transition. The corresponding onset threshold $\delta_{\rm c}$, which for $x\!=\!1/2$ takes the value $\delta_{\rm c}\!\approx\!0.49$, is marked in all three panels by the vertical dashed line. The black squares pertain to the disordered-crystal ensembles that we study next, for which $\delta\!<\!\delta_{\rm c}$.

To further substantiate the amorphization transition at $\delta_{\rm c}$ and to establish the existence of finite long-range order in our disordered crystals for $\delta\!<\!\delta_{\rm c}$, we consider the pair correlation function $g(r)$ among the `large' particle-species; results for $\delta\!=\!0.38,0.44,0.49,0.70$ are shown in Fig.~\ref{fig:fig2}. We find that solids with $\delta\!\lesssim\!\delta_{\rm c}$ all feature finite long-range order, indicated by the presence of peaks in $g(r)$ that persist to the largest distances $r$ considered. In particular, long-range order is clearly present in the ensembles pertaining to $\delta\!=\!0.38$ to $\delta\!=\!0.44$, which are studied next.

\section{Universal nonphononic VDoS}

In order to address the emergence of QLEs and their possible universal nonphononic VDoS in disordered crystals, it is important to select sufficiently small systems such that a phonon-free frequency window opens below the first shear-wave frequency, as explained at length in~\cite{JCP_Perspective}. Furthermore, since QLEs that emerge in structural glasses are characterized by a core size $\xi_{\rm g}$ --- usually on the order of a few interparticle distances~\cite{JCP_Perspective,pinching_pnas} --- it is also important that the linear system size is chosen to be sufficiently larger than $\xi_{\rm g}$, in order for QLEs to fit comfortably in the simulation box~\cite{JCP_Perspective,lerner2019finite}. To accommodate these two requirements, we opt for studying emergent QLEs in disordered crystals of $N\!=\!1372$ particles. We prepared 667900, 297600, 150000, and 109600 independent realizations of disordered crystals with $\delta\!=\!0.38,0.40,0.42$ and $0.44$ (all below $\delta_{\rm c}$), following the protocol described above and in Appendix~\ref{app:model}. For each disordered crystal, we calculated the first 30 nontrivial vibrational modes $\psiv_\omega$ as defined via Eq.~(\ref{eigenvalue_equation}).

The resulting VDoS are shown in Fig.~\ref{fig:fig3}a, which presents a key result of this work. We find that disordered crystals featuring finite long-range order host a population of quasilocalized modes whose frequencies obey the universal $\sim\!\omega^4$ law. Having demonstrated the validity of the nonphononic quartic scaling law in disordered crystals, we turn now to discussing the non-universal prefactor $A_{\rm g}$ in the full expression of the VDoS, ${\cal D}(\omega)\!=\!A_{\rm g}\,\omega^4$. The prefactor $A_{\rm g}$, of dimension $[\mbox{time}]^5$, has been suggested to serve as an indicator of the abundance of soft QLEs~\cite{modes_prl_2016,cge_paper}. As such, it was the focus of several recent studies~\cite{pinching_pnas,david_fracture_2021,sticky_spheres1_karina_pre2021,phonon_widths2,incompressible_letter_2022}; notably, $A_{\rm g}$ was shown to strongly correlate with the tensile fracture toughness of glassy solids~\cite{david_fracture_2021}. The prefactors $A_{\rm g}$ of our disordered crystals --- obtained here by fits (represented by the straight lines) to a $\sim\!\omega^4$ power-law at low frequencies --- are reported in Fig.~\ref{fig:fig3}b. We find an approximate exponential dependence of $A_{\rm g}$ on the parameter $\delta$, which is reminiscent of the Boltzmann-like dependence of $A_{\rm g}$ on the \emph{equilibrium} parent temperature of structural glasses quenched from deeply supercooled liquids~\cite{pinching_pnas,mw_thermal_origin_of_qle_pre2020}. A similar exponential decay of $A_{\rm g}$ was also observed recently in a mean-field spin-glass model~\cite{scipost_mean_field_qles_2021}.

\section{QLE properties}

We next study some of the structural and mechanical properties of the low-frequency QLEs that emerge in our disordered crystals. To this aim, we invoke the nonlinear framework of~\cite{SciPost2016,episode_1_2020}, which provides a robust representation of soft QLEs, denoted as $\piv$ (see also Appendix~\ref{app:observables}). We first focus on the spatial decay of QLEs, which is presented in the inset of Fig.~\ref{fig:fig3}a for two randomly selected nonlinear modes. It is observed that similarly to QLEs in structural glasses~\cite{JCP_Perspective}, QLEs in our disordered crystals feature a disordered core of a few interparticle distances, followed a power-law $\sim\!r^{-2}$ tail (i.e.~$\sim\!r^{-\dbar+1}$ in three dimensions, $\dbar\!=\!3$).

The degree of spatial localization of QLEs is further discussed next, by considering the participation ratio $e$, defined as $e(\piv)\!\equiv\!N^{-1}\big(\sum_i \piv_i\cdot\piv_i\big)^2\!/\!\sum_i\big(\piv_i\cdot\piv_i\big)^2$. Here, $\piv_i$ denotes the $\dbar$-dimensional vector of Cartesian components pertaining to the $i^{\mbox{\tiny th}}$ particle of a QLE $\piv$. The participation ratio $e(\piv)$ quantifies the degree of spatial localization of $\piv$; in particular, $\piv$'s that are localized on a core of $N_{\rm c}$ particles are expected to follow $Ne\!\approx\!N_{\rm c}$, whereas spatially extended $\piv$ are expected to follow $Ne\!\sim\!N$ (or $e\!\sim\!1$). In Fig.~\ref{fig:fig3}c, we plot the factored means $N\bar{e}$ vs.~$\delta$; here, $\bar{e}$ was calculated over 1000 QLEs (for each value of $\delta$) in systems of $N\!=\!16384$. Also shown are the 2nd and 3rd quartiles of the products $Ne$ of the QLEs, which are covered by the error bars. We find that moving away from the amorphization transition, giving rise to less disordered states, leads to a stronger localization of QLEs. This enhanced localization is reminiscent of the stronger localization of QLEs observed in structural glasses with increasing glass stability, as shown in e.g.~\cite{pinching_pnas,LB_modes_2019}.

\begin{figure}[ht]
\centering
\begin{tabular}{ccc}
\includegraphics[width=1.0\linewidth]{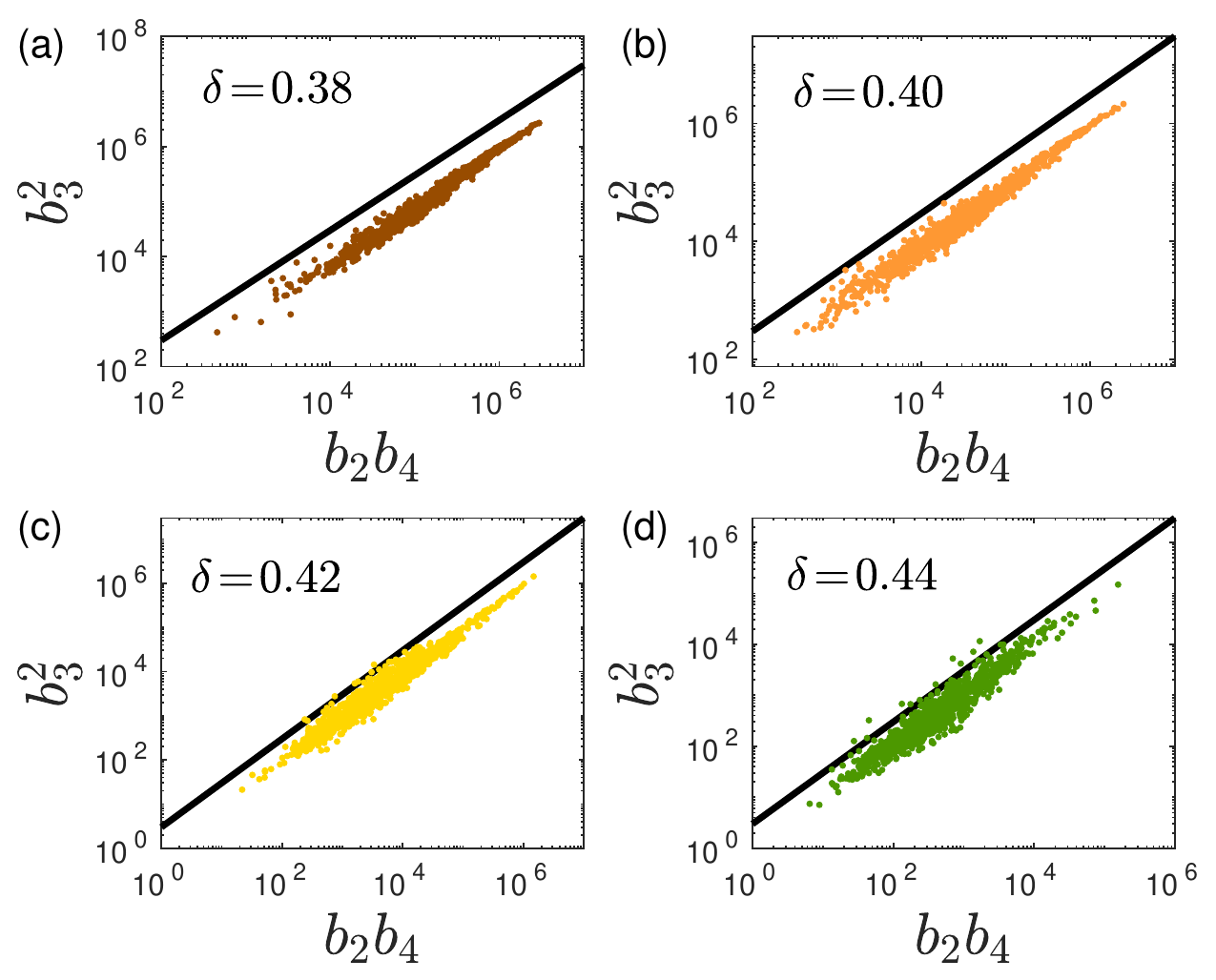}
\end{tabular}
\caption{The square of the cubic Taylor series coefficient of QLEs, $b_3^2$, is scatter-plotted against the product of the corresponding quadratic and quartic expansion coefficients $b_2b_4$, see text for further details. The continuous lines represent the stability bound $b_3^2\!=\!3b_2b_4$. The emerging picture, and its relation to the corresponding picture in structural glasses quenched from a melt~\cite{episode_1_2020}, are discussed in the text.}
\label{fig:b2b3b4}
\end{figure}

We next turn to the micromechanical properties of QLEs. In particular, for each QLE $\piv$ we calculate the first 3 nonzero Taylor series coefficients --- denoted $b_2,b_3,b_4$ respectively --- of the energy difference $U(s)\!-\!U(0)$, assuming particles are displaced a distance $s$ along the generalized direction defined by $\piv$ about the local minimum of the potential energy $U$. In Fig.~\ref{fig:b2b3b4}, we scatter-plot $b_2^2$ vs.~the products $b_2b_4$ for QLEs of different $\delta$-ensembles; from stability considerations, one expects $b_3^2\!\le\!3b_2b_4$, which means that the system sits in a single-well local potential, or in the stable well amongst a double-well local potential, see further discussions e.g.~in~\cite{episode_1_2020,JCP_Perspective}.

\begin{figure}[ht]
\centering
\begin{tabular}{ccc}
\includegraphics[width=1.0\linewidth]{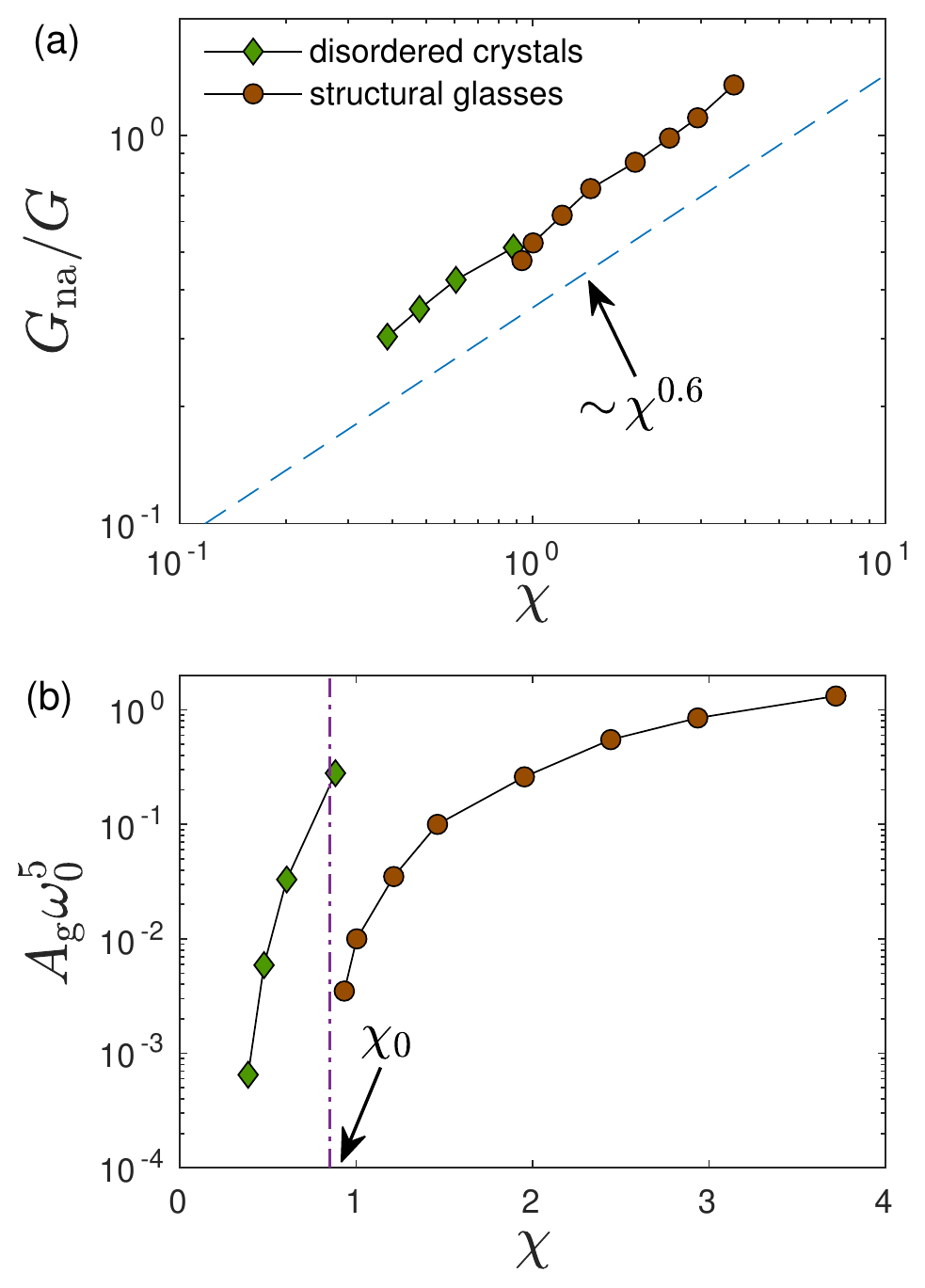}
\end{tabular}
\caption{Relations between quantifiers of mechanical disorder. (a) $G_{\rm na}/G$ vs.~$\chi$ for disordered crystals and structural glasses, as indicated by the legend. The two data sets appear to approximately form a single function, see text for further discussion. (b) The dimensionless nonphononic VDoS prefactor $A_{\rm g}\omega_0^5$ vs.~$\chi$ for disordered crystals and structural glasses (the same symbols as in panel (a)). In sharp constract to the results of panel (a), we find that the two classes of disordered solids follow very different curves. In particular, the disordered crystals appear to have anomalously low $\chi$ values relative to the number of QLEs they host.}
\label{fig:quantifiers}
\end{figure}

We find that QLEs in disordered crystals with $\delta\!=\!0.38$ typically reside quite below the stability bound $b_3^2\!=\!3b_2b_4$ (represented by the continuous lines in Fig.~\ref{fig:b2b3b4}), but less so as $\delta$ is increased. For $\delta\!=\!0.44$, we find that in fact a small fraction of QLEs violate the stability bound. This scenario is reminiscent of the behavior of QLEs observed in structural glasses with various degrees of stability, as shown in~\cite{episode_1_2020}. There, glasses quenched from very deeply supercooled states featured QLEs for which typically $b_3^2\!<\!3b_2b_4$, while glasses made by quenching high-energy liquid states featured a sizeable fraction QLEs that violate this stability bound. Also similar to the phenomenology of structural glasses is the appearance of a lower bound of the form $b_3^2\!\propto\!b_2b_4$, but with a smaller prefactor compared to the aforementioned stability bound. These two bounds suggest the existence of a deep relation between excitations' frequencies and their associated cubic anharmonicity, as discussed at length in~\cite{episode_1_2020,JCP_Perspective}.

\section{Mechanical-disorder quantifiers}

Having characterized the emergent nonphononic VDoS in disordered crystals and the properties of the underlying excitations, we turn now to examining how dimensionless quantifiers of mechanical disorder relate to each other in disordered crystals, and how those compare to the corresponding relations in structural glasses. For example, a scaling relation between $A_{\rm g}$ and the strength of spatial fluctuation of the shear modulus quantified by $\chi$ has been demonstrated in computer structural glasses~\cite{david_fracture_2021,phonon_widths2,incompressible_letter_2022}. The generality of such relations between different measures of mechanical disorder remains largely unexplored across various classes of disordered solids. In order to shed light on this important issue, we employ --- in addition to our disordered crystals dataset --- a dataset of a simple glass-forming model (see Appendix~\ref{app:structural_glass} for details) that can be supercooled down to extremely low temperatures using the Swap Monto-Carlo algorithm~\cite{LB_swap_prx}, spanning a broad range of mechanical disorder.

Our results for the interrelations between the dimensionless quantifiers of mechanical disorder $G_{\rm na}/G$, $\chi$ and $A_{\rm g}$ are presented in Fig.~\ref{fig:quantifiers}. In Fig.~\ref{fig:quantifiers}a, we plot $G_{\rm na}/G$ vs.~$\chi$ for both disordered crystals and structural glasses. Interestingly, while the data are somewhat noisy, both datasets for these two systems appear to lie on the {\em same} approximate power-law $G_{\rm na}/G\!\sim\!\chi^{0.6}$. This remarkable commonality of the two classes of disordered solids suggests that the two quantifiers in both classes are sensitive to mechanical disorder in a similar fashion.

The situation is dramatically different, however, when comparing the relation between the prefactor $A_{\rm g}$ --- made dimensionless by multiplying it by $\omega_0^5$ --- and $\chi$, considered in both disordered crystals and structural glasses, as done in Fig.~\ref{fig:quantifiers}b. Here, we find that these two classes of disordered solids exhibit very different $A_{\rm g}\omega_0^5$--$\chi$ relations. Most notably, for similar values of the dimensionless prefactor $A_g\omega_0^5$ --- which encodes information about both the number and characteristic frequency of QLEs~\cite{cge_paper,pinching_pnas} ---, $\chi$ of disordered crystals is significantly smaller. Put differently, we observe that disordered crystals appear to feature many more QLEs than  expected from the $A_g\omega_0^5$--$\chi$ relation of structural glasses. Finally, we note that in a recent work it was found that $\chi$ of a broad variety of (computer) structural glasses appears to be bounded from below by $\chi_0\!\approx\!0.85$~\cite{karina_minimal_disorder}, marked by the vertical, dash-dotted line in Fig.~\ref{fig:quantifiers}b. Interestingly, in our disordered crystals, for the range of $\delta$ studied here, $\chi$ breaks this apparent bound by a significant factor.

\section{Discussion and outlook}

In this work, we theoretically studied the emergence of glass-like properties in disordered crystals, most notably the emergence of low-energy, quasilocalized excitations (QLEs), whose existence has been recently established in structural glasses quenched from a melt~\cite{modes_prl_2016,modes_prl_2018,modes_prl_2020,JCP_Perspective}. By continuously varying the degree of mechanical disorder in disordered crystals in a model that features an amorphization transition, we found that {\em below} the transition --- where the disordered crystals feature finite long-range order --- QLEs emerge and their frequencies follow the universal $\sim\!\omega^4$ nonphononic VDoS as seen in many classes of structural glasses~\cite{modes_prl_2016,modes_prl_2018,modes_prl_2020,JCP_Perspective} and other models~\cite{parisi_spin_glass,scipost_mean_field_qles_2021,meanfield_qle_pierfrancesco_prb_2021,vector_spin_glasses_scipost_2022,Chalker2003,Gurevich2003,itamar_gps_prb2020}. As such, our results extend the degree of universality of the $\omega^4$ law to include disordered crystals.

We showed that the non-universal prefactor $A_{\rm g}$ of the universal $\sim\!\omega^4$ VDoS, which is related to the number of QLEs~\cite{pinching_pnas}, varies substantially with a control parameter of the model, similarly to an apparently related variation observed in structural glasses~\cite{pinching_pnas,LB_modes_2019}. We further found that the micromechanical properties of QLEs in disordered crystals, e.g.~their spatial localization properties, resemble those observed in structural glasses, and that they satisfy a similar stability bound. Finally, considering interrelations between different quantifiers of mechanical disorder in both disordered crystals and structural glasses, striking similarities and differences emerged. We discovered that the relation between a measure of elastic nonaffinity, $G_{\rm na}/G$ (cf.~\eqref{eq:G_na}), and a measure of shear modulus fluctuations, $\chi$ (cf.~\eqref{eq:chi}), appears to form a unique function for both disordered crystals and structural glasses. On the other hand, not only disordered crystals feature $\chi$ values lower than the lowest value ever obtained in structural glasses~\cite{karina_minimal_disorder}, but their $A_{\rm g}(\chi)$ relation also implies that they host many more QLEs than
the corresponding structural glasses of the same $\chi$.

The extended validity of the universal glassy $\sim\!\omega^4$ VDoS that includes also disordered crystals, as established in this work, appears to be intimately related to known glass-like anomalies experimentally observed in laboratory disordered crystals at cryogenic temperatures~\cite{ackerman1981glassy,Anderson_1985, Pohl_disordered_crystals_1992,ramos1997quantitative,pohl_review,ramos2003low,vdovichenko2015thermal,gebbia2017glassy,moratalla2019emergence,Ramos_2020}. Most notably, our findings regarding the low-frequency $\sim\!\omega^4$ VDoS appear to be related to the existence of a $\sim\!T^5$ contribution to the specific heat at low temperatures $T$ in disordered crystals~\cite{ramos2003low,vdovichenko2015thermal,gebbia2017glassy,moratalla2019emergence,Ramos_2020}, in addition to a $\sim\!T$ contribution associated with two-level systems~\cite{Zeller_and_Pohl_prb_1971,anderson1972anomalous,phillips1972tunneling} and Debye's $\sim\!T^3$ contribution of phonons~\cite{kittel2005introduction}. They may also be related to low-temperature anomalies in the heat conductivity~\cite{pohl_review} and vibrational anomalies, such as the emergence of a Boson peak~\cite{moratalla2019emergence}, documented in disordered crystals. Overall, our findings support the experimentally-inferred suggestion that the absence of long-range order is neither sufficient nor necessary for the existence of the
low-energy excitations~\cite{pohl_review,ramos1997quantitative}.

In order to explore in more quantitative terms the relations between the model studied in this work and laboratory disordered crystals, future work should more realistically represent the interactions between different species (e.g.~between impurity and host atoms, captured here by the parameter $\delta$) and vary the species concentration $x$ (fixed here at $x\!=\!1/2$), which is the key control parameter in laboratory binary disordered crystals. In particular, as the prefactor of the $\sim\!T^5$ contribution to the specific heat would be proportional to the prefactor $A_{\rm g}$ of the $\sim\!\omega^4$ VDoS, such future studies would need to resolve $A_{\rm g}(\delta,x)$ (the results for $A_{\rm g}(\delta,x\!=\!1/2)$, for $\delta$ defined in~\eqref{eq:delta}, are shown in Fig.~\ref{fig:fig3}b). Likewise, extensions to multi-component disordered crystals can be explored.

Our results also raise basic questions about disordered solids in a broader context. One class of such questions is concerned with the identification of the essential ingredients necessary for the emergence of universal nonphononic excitations in disordered solids. In this context, we note that our disordered crystals results appear to echo the theoretical framework recently developed in~\cite{avraham_minimal_complexes_2021}. There, it was shown that the self-consistent introduction of positional disorder and internal stresses (generated also in our model) into small groups of particles that feature perfect crystalline order --- termed ``minimal complexes'' --- may lead to the emergence of the $\sim\!\omega^4$ VDoS~\cite{avraham_minimal_complexes_2021}. Future work should clarify whether indeed such a connection between our findings and those of~\cite{avraham_minimal_complexes_2021} exist, and if so, explore its possible implications. It would be also interesting to explore the relations between our findings and the model of~\cite{Gurevich2003}, which a priori assumes the existence of soft quasilocalized excitations. Finally, another interesting question is concerned with the origin and degree of universality of the approximate relation $G_{\rm na}/G\!\sim\!\chi^{0.6}$, observed in Fig.~\ref{fig:quantifiers}a to hold in both disordered crystals and structural glasses, which is yet another topic for future investigation.

\acknowledgements
We warmly thank Geert Kapteijns and Jeppe Dyre for discussions that led to this work. E.L.~acknowledges support from the NWO (Vidi grant no.~680-47-554/3259). E.B.~acknowledges support from the Ben May Center for Chemical Theory and Computation and the Harold Perlman Family.

\appendix

\section{Disordered crystals model}
\label{app:model}

We employ a model of $N$ equal-mass, point-like particles in three dimensions, which derives from the model studied in~\cite{barrat_2013_amorphization,barrat_2014_pnas_amorphization,barrat_2016_prb_amorphization}. A fraction $x\!=\!1/2$ of the particles are randomly selected and labelled, to be effectively ``inflated'' in order to introduce quenched microscopic disorder into the system, as explained below. Pairs of particles interact via a Lennard-Jones pairwise potential of the form
\begin{equation}
    \varphi_{\mbox{\tiny LJ}}(r_{ij},\sigma_{ij})=4\varepsilon\bigg[ \big(\sFrac{\sigma_{ij}}{r_{ij}}\big)^{12} - \big(\sFrac{\sigma_{ij}}{r_{ij}}\big)^6 + \sum_{\ell=0}^2c_{2\ell}\big(\sFrac{r_{ij}}{\sigma_{ij}}\big)^{2\ell}\bigg] \,,
\end{equation}
for $r_{ij}/\sigma_{ij}\!<\!2.5$, and $\varphi_{\mbox{\tiny LJ}}(r_{ij},\sigma_{ij})\!=\!0$ otherwise. Here $\varepsilon$ sets the microscopic units of energy, and the length parameters $\sigma_{ij}$ are set as
\begin{equation}
\label{eq:delta}
\sigma_{ij} = \left\{
\begin{array}{cc}
\lambda_{\mbox{\tiny SS}} & \mbox{both $i,j$ are unlabelled} \\
\delta (\lambda_{\mbox{\tiny SL}}-\lambda_{\mbox{\tiny SS}}) + \lambda_{\mbox{\tiny SS}}& \mbox{either $i$ or $j$ are labelled}\\
\delta (\lambda_{\mbox{\tiny LL}}-\lambda_{\mbox{\tiny SS}}) + \lambda_{\mbox{\tiny SS}} & \mbox{both $i$ and $j$ are labelled}
\end{array}
\right.\,.
\end{equation}
We chose the length parameter $\lambda_{\mbox{\tiny SS}}$ as the microscopic unit of length, and set $\lambda_{\mbox{\tiny SL}}\!=\!1.18\lambda_{\mbox{\tiny SS}}$ and $\lambda_{\mbox{\tiny LL}}\!=\!1.4\lambda_{\mbox{\tiny SS}}$. The parameter $\delta$, which controls the effective interaction of labelled particles, determines the degree of microscopic disorder: $\delta\!=\!0$ corresponds to a mono-component system, whereas $\delta\!=\!1$ corresponds to a 50:50 binary mixture of `large' and `small' particles of 1.4:1 ratio size, as used in popular glass-forming models. The coefficients $c_4\!=\!\mbox{6.201261686784e-04}$, $c_2\!=\!\mbox{-9.70155098112e-03}$ and $c_0\!=\!\mbox{4.0490237952e-02}$ are determined such that the potential is continuous up to two derivatives at the dimensionless cutoff $r^c_{ij}/\sigma_{ij}\!=\!2.5$.

We construct disordered crystals by setting $\delta\!=\!0$ and placing our $N$ particles on a fcc lattice. Then, we increment the parameter $\delta$ and minimize the potential energy by a standard nonlinear conjugate-gradient method, combined with a Berendsen-like barostat~\cite{berendsen,allen1989computer} fixing the pressure at zero. This minimization step introduces positional disorder and internal stresses into the system. For $\delta\!<\!0.35$, we use increments of $0.05$, whereas for higher $\delta$'s the increments are set to $0.01$. We studied a wide range of system sizes, from $N\!=\!1372$ to $N\!=\!108000$.

\section{Observables}
\label{app:observables}

The observables we consider are detailed next; we consider vibrational modes' frequencies $\omega$ obtained via the eigenvalue equation
\begin{equation}
\label{eigenvalue_equation}
    \calBold{H}\cdot\psiv_{\omega} = \omega^2\psiv_\omega\,,
\end{equation}
where $\calBold{H}\!=\!\frac{\partial^2U}{\partial\xv\partial\xv}$ with $U\!=\!\sum_{i<j}\varphi_{\mbox{\tiny LJ}}(r_{ij},\sigma_{ij})$ being the potential energy and $\xv$ the particles' spatial coordinates. $\psiv_\omega$ is the eigenvector pertaining to the eigenvalue $\omega^2$, and we note that all masses are set to unity. The harmonic (linear) normal modes $\psiv_\omega$ are used to determine the VDoS.

The properties of QLEs are studied using the nonlinear framework discussed e.g.~in~\cite{SciPost2016,episode_1_2020}. Within this framework, QLEs are solutions $\piv$ to the equation
\begin{equation}\label{eq:cubic_modes}
    \calBold{H}\cdot\piv = \frac{\calBold{H}:\piv\piv}{{\bm U'''}\tripleCdot\piv\piv\piv}{\bm U'''}:\piv\piv\,,
\end{equation}
where ${\bm U'''}\!\equiv\!\frac{\partial U}{\partial\xv\partial\xv\partial\xv}$ denotes the rank-3 tensor of derivatives of the potential energy, and $\cdot,:,\tripleCdot$ denote single, double or triple contractions over particle indices and Cartesian components, respectively. In~\cite{SciPost2016,episode_1_2020}, it was shown that QLEs defined via Eq.~(\ref{eq:cubic_modes}) closely resemble quasilocalized vibrational (harmonic) modes, in the absence of hybridizations with low-frequency phonons. Low-energy QLEs were calculated as explained in Appendix~B of~\cite{episode_1_2020}, which presumably provides the single, lowest-energy QLE in a given disordered solid or glass sample. The spatial decay of QLEs was calculated following~\cite{modes_prl_2016}.

We considered the athermal shear and bulk moduli, denoted $G$ and $K$ respectively, whose precise definitions can be found e.g.~in Appendix~A of~\cite{sticky_spheres1_karina_pre2021}. We also considered the widely studied Poisson's ratio $\nu$, given by
\begin{equation}
    \nu = \frac{3K-2G}{6K+2G}\,.
\end{equation}

In order to further quantify mechanical disorder, we consider the sample-to-sample relative fluctuations of the shear modulus, namely
\begin{equation}
\label{eq:chi}
    \chi \equiv \frac{\sqrt{N\overline{(G - \overline{G})^2}}}{\overline{G}}\,,
\end{equation}
where $\overline{\bullet}$ denotes the sample-to-sample average. The quantifier $\chi$ was discussed at length in~\cite{phonon_widths,scattering_letter_jcp_2021,phonon_widths2,sticky_spheres1_karina_pre2021,massimo_prl_2021_scattering}; it has been shown to determine wave attenuation rates in the harmonic regime~\cite{scattering_letter_jcp_2021,massimo_prl_2021_scattering}. Here, we use a jackknife-like method to approximate $\chi$, see details in~\cite{phonon_widths2}.

Finally, we considered another dimensionless quantifier of mechanical disorder defined as the ratio between what is known as the {\it nonaffine} term of the shear modulus --- denoted $G_{\rm na}$ --- to the total shear modulus $G$, namely
\begin{equation}
\label{eq:G_na}
    \frac{G_{\rm na}}{G} = \frac{\frac{\partial^{2}U}{\partial \gamma\partial \xv} \cdot \calBold{H}^{-1}\cdot \frac{\partial ^{2}U}{\partial\xv\partial \gamma}}{VG}\,,
\end{equation}
where $\gamma$ is the shear-strain parameter as defined e.g.~in~\cite{sticky_spheres1_karina_pre2021}.

\section{Structural glass model}
\label{app:structural_glass}

In order to compare the behavior of our disordered crystals to structural glasses formed by quenching a melt, we employed the dataset of soft-sphere structural glasses quenched from a broad range of equilibrium temperatures $T_{\rm p}$ as described in~\cite{pinching_pnas,phonon_widths2,boring_paper}. This model can be equilibrated down to very low temperatures using the Swap-Monte-Carlo method~\cite{LB_swap_prx}. Our glass ensembles consist of 2000 independent samples of $N\!=\!16384$ particles each.

%

\end{document}